\documentclass[a4paper,11pt]{article}
\usepackage{pos}

\title{Features in the Cosmic Ray Energy Spectrum Observed with Telescope Array Surface Detectors}
\ShortTitle{TA SD Spectrum Features}

\author*[a]{Jihyun Kim}
\author[a]{Dmitri Ivanov}
\author[a]{Gordon Thomson}

\affiliation[a]{High Energy Astrophysics Institute and Department of Physics and Astronomy, University of Utah\\
Salt Lake City, Utah 84112-0830, USA}

\onbehalf{on behalf of the Telescope Array Collaboration} 

\emailAdd{jihyun@cosmic.utah.edu}

\abstract{Ultra-high energy cosmic rays (UHECRs) are extremely energetic charged particles that originate from outer space. The Telescope Array (TA) experiment, the largest UHECR observatory in the Northern Hemisphere, has provided high-precision measurements of the cosmic ray energy spectrum due to its stable operation and efficient data collection. These measurements have revealed three significant spectral features: the ankle, shoulder, and cutoff. Analyzing these features is crucial for understanding the origin and propagation of UHECRs. In this talk, we will present the latest energy spectrum measured by the TA surface detectors and discuss the observed differences in the UHECR energy spectrum between the northern and southern skies.}

\ConferenceLogo{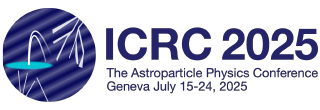}

\FullConference{39th International Cosmic Ray Conference (ICRC2025)\\
 15–24 July 2025\\
Geneva, Switzerland\\}

\begin{document}
\maketitle

\vspace{-3mm}
\section{Introduction}
\vspace{-3mm}
Ultra-high energy cosmic rays (UHECRs), which are cosmic rays with energies greater than $10^{18}$~eV, are among the most energetic charged particles known to reach Earth from outer space. These particles serve as powerful probes of the universe, offering valuable insights into extreme astrophysical phenomena. Research on UHECRs typically focuses on three key observables: the energy spectrum, mass composition, and arrival directions. It is essential to analyze these aspects to advance our understanding of the origins, nature, and propagation mechanisms of UHECRs from their sources across the universe.

The energy spectrum of UHECRs provides critical information for constraining models of their astrophysical sources and propagation mechanisms. For example, a significant spectral feature is the suppression at energies $\sim$$10^{19.8}$~eV, at the energy of the GZK, as predicted by Greisen, Zatsepin, and Kuzmin~\cite{Greisen-1966-PRL-16-748, Zatsepin-1966-JETPL-4-78}. This feature arises due to interactions between UHECR protons and photons from the cosmic microwave background, which result in photo-pion production that restricts the distance these particles can travel. 

The High Resolution Fly’s Eye (HiRes) experiment first reported observational evidence of this cutoff~\cite{HiRes:2007lra}, which was later confirmed by the Telescope Array (TA)~\cite{TelescopeArray:2012qqu}. In the southern hemisphere, the Pierre Auger \textcolor{black}{collaboration} (Auger) also observed a comparable flux suppression, though at slightly lower energies~\cite{PierreAuger:2008rol}. While the GZK mechanism remains a leading explanation for the HiRes observation, the high-energy cutoff may also reflect the maximum energy achievable by UHECR sources or result from other propagation effects. Determining the dominant mechanism behind the cutoff is an ongoing challenge.

This study examines the features of the cosmic ray energy spectrum using data collected by the Telescope Array surface detector array over a 16-year period from May 11, 2008, to May 10, 2024. Our goal is to enhance understanding of the UHECR energy spectrum and identify the physical mechanisms responsible for its structural characteristics.

\vspace{-3mm}
\section{Telescope Array}
\vspace{-3mm}
The Telescope Array (TA), located in the western Utah desert, USA, at coordinates $39.3^\circ \mathrm{N}$, $112.9^\circ \mathrm{W}$, is the largest UHECR observatory in the Northern Hemisphere. It is designed to observe extensive air showers---cascades of millions of subatomic particles initiated when a single UHECR collides with a nucleus in Earth's atmosphere. The TA is strategically positioned at an elevation of 1,400 meters above sea level to capture extensive air showers at their maximum development. It operates as a hybrid detector, combining a surface detector array and air-fluorescence detectors. The surface detector (SD) array comprises 507 scintillation counters arranged in a grid with 1.2 km spacing, spanning an area of approximately 700 km$^2$. Each SD counter contains two layers of plastic scintillators to record particle footprints as extensive air showers reach the ground~\cite{TelescopeArray:2012uws}. Additionally, three fluorescence detector (FD) stations, equipped with 38 telescopes, monitor the sky above the SD array, covering an elevation range of $3^\circ$--$31^\circ$. These telescopes capture ultraviolet emissions produced as extensive air showers propagate through the atmosphere~\cite{Tokuno:2012mi}.

\vspace{-3mm}
\section{Surface Detector Event Reconstruction}
\vspace{-3mm}

\begin{figure}
    \centering
    \includegraphics[width=1\linewidth]{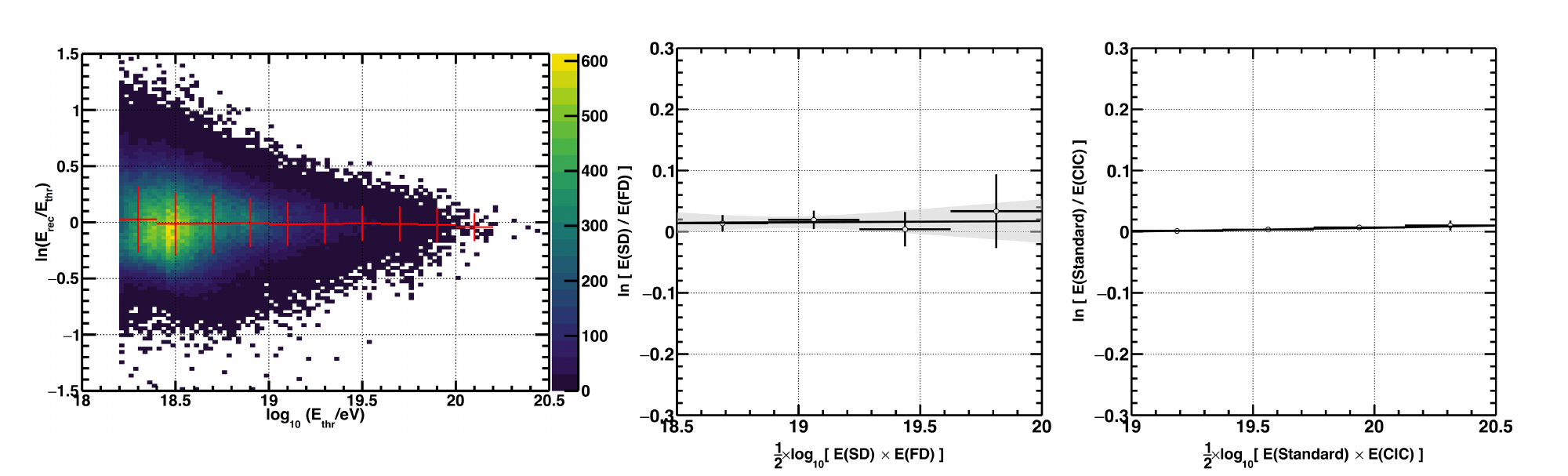}
    \caption{\textbf{Linearity in energy reconstruction.} The ratios of reconstructed to thrown energy in the MC simulation (left), reconstructed energy measured by the SD and FD (middle), and the TA standard energy reconstruction to the CIC method (right), all as functions of energy, consistently demonstrate the linearity of the energy reconstruction.}
    \label{fig:enLinearity}
    \vspace{-3mm}
\end{figure}


The reconstruction of extensive air showers detected by the SD array involves a multi-step procedure. Initially, the shower core location and the arrival direction of the primary cosmic ray are determined using spatial and timing information from the triggered SD counters. A modified Linsley shower-shape function fit is used to determine these properties~\cite{Teshima:1986rq}. Subsequently, the lateral distribution of secondary particles is modeled using the same functional form adopted by the AGASA experiment~\cite{Yoshida:1994jf, Takeda:2002at}. From this fit, we interpolate the particle density at a reference distance of 800 meters from the shower axis, $S(800)$, which serves as an energy estimator. This reference distance is optimized for the TA site's altitude and detector spacing, while also minimizing systematic uncertainties associated with different primary particle types.

The next step in the SD energy reconstruction is based on a Monte Carlo (MC) study, supported by detailed data/MC comparisons, resulting in an energy reconstruction table, \textcolor{black}{commonly referred to as a lookup table}~\cite{TelescopeArray:2014nxa}. To estimate the primary cosmic ray energy, we use a high-statistics SD MC simulation generated with the CORSIKA framework~\cite{1998cmcc.book.....H}, incorporating the QGSJET-II-03 hadronic interaction model~\cite{Ostapchenko:2004ss}. \textcolor{black}{The simulation assumes proton primaries, based on mass composition measurements by HiRes and the TA~\cite{HiRes_composition, TA_hybrid_composition, TA_SD_composition}.} The initial energy estimate, $E_{\rm TBL}$, is derived from $S(800)$ and the reconstructed zenith angle, $\theta$, via $\sec(\theta)$. To reduce model-dependent energy scale uncertainty, this estimate is calibrated against calorimetric energy measurements from the fluorescence detector (FD) using hybrid events observed by both SD and FD simultaneously. A scaling factor of 1.27 is obtained from this comparison, leading to the final energy estimate: $E_{\rm Final} = E_{\rm TBL}/1.27$. 

\textcolor{black}{We evaluate the systematic uncertainties associated with the linearity of the SD energy reconstruction using three approaches, shown in Figure~\ref{fig:enLinearity}. The left panel displays the ratio of reconstructed to thrown energy in MC simulations, using the same reconstruction program \textcolor{black}{for the MC as is} applied in data analysis. This comparison confirms the linearity of the reconstruction. Additionally, we incorporate extended hybrid data and re-examine the ratio of SD to FD energies as a function of energy. The middle panel presents this ratio along with a linear fit, confirming the absence of significant energy-dependent non-linearity.}

To further evaluate the systematic uncertainties, we employ an alternative data-driven approach: the constant intensity cut (CIC) method~\cite{Hersil:1961zz}. This method, also used by the Pierre Auger Observatory (Auger), provides a valuable cross-check of the MC-based energy reconstruction and serves as an independent benchmark for validation.

The CIC energy assignment is performed in two stages. First, the measured $S(800)$ is de-attenuated to $S_{34}$, representing the equivalent signal the shower would produce at the reference zenith angle of $34^\circ$, which corresponds to the average zenith angle of \textcolor{black}{events observed by} TA. In the second stage, $S_{34}$ is converted to energy using a power law relationship derived from comparisons with calorimetric energy measurements from the FD. To account for uncertainties in both FD and SD measurements, an iterative procedure is applied. This involves comparing $S_{34}$ to the geometric mean of the FD energy and the CIC-derived energy, repeating the process until convergence is achieved within the fitting uncertainties.

The resulting energy spectra are consistent with those obtained from the MC-based method within a 2\% uncertainty (see Figure 4 in~\cite{TelescopeArray:2023bdy}). \textcolor{black}{The right panel of Figure~\ref{fig:enLinearity} presents the ratio of standard to CIC-derived energies as a function of energy, again showing no significant energy-dependent difference.} Since the CIC method is entirely data-driven and does not rely on MC simulations, this confirms the robustness of the energy reconstruction of the standard MC-based method.

\vspace{-3mm}
\section{Features in the Cosmic Ray Energy Spectrum}
\label{sec:spec_features}
\vspace{-3mm}

\begin{figure}
    \centering
    \includegraphics[width=0.55\linewidth]{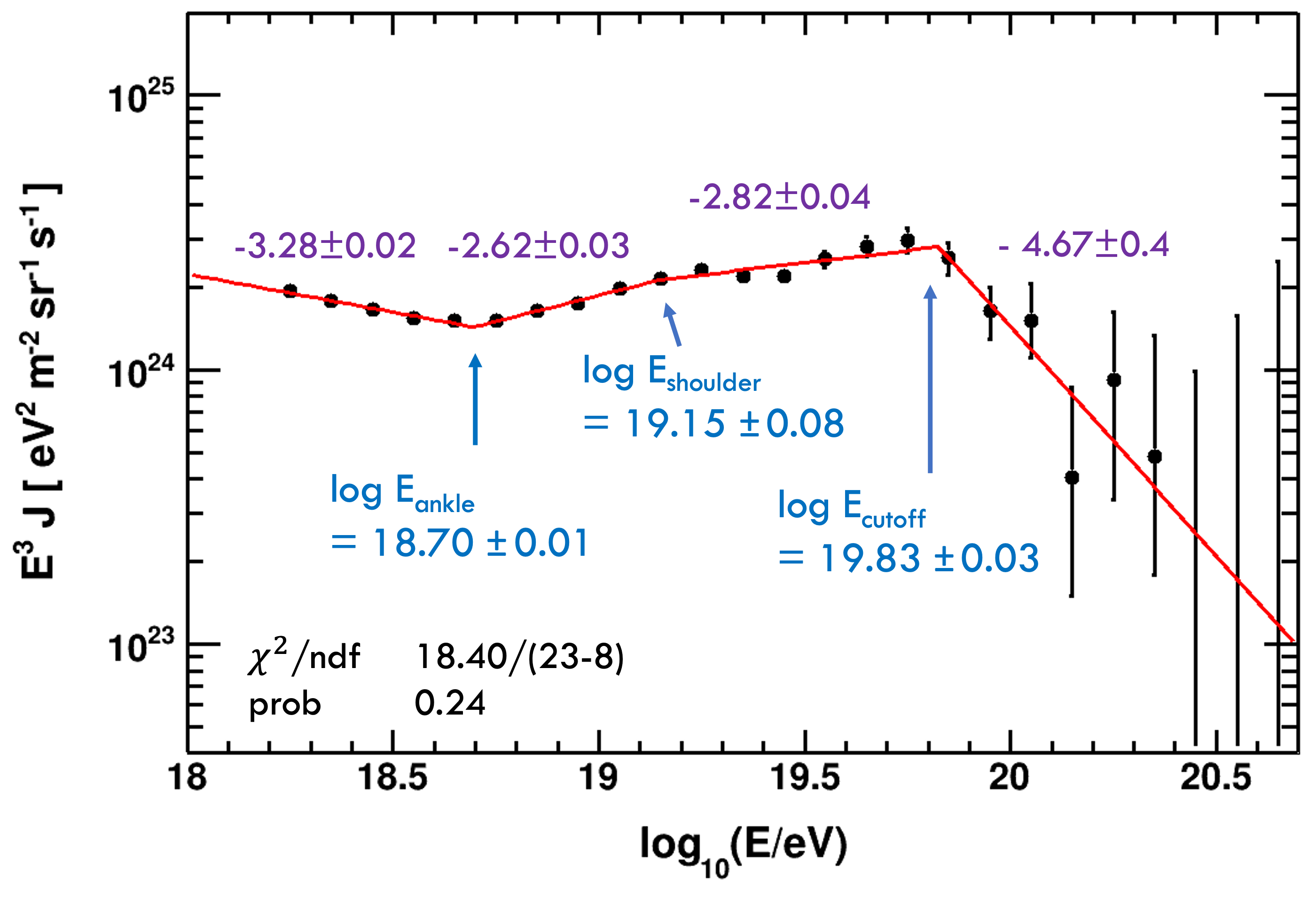}
    \caption{\textbf{Energy spectrum from 16 years of TA SD observations.} Black markers with error bars indicate the measured data, while the red line represents a fit using a thrice-broken power law model. This model incorporates three energy breakpoints ($E_{\mathrm{ankle}}$, $E_{\mathrm{shoulder}}$, and $E_{\mathrm{cutoff}}$) and four spectral slopes ($p_1$, $p_2$, $p_3$, and $p_4$). The resulting fit parameters are superimposed on the graph.}
    \label{fig:spectrum_features}
\end{figure}

This section presents the spectral features observed in the UHECR flux, as measured by the TA SD array over a 16-year period, from May 11, 2008, to May 10, 2024. The analysis applies the following event selection criteria described in~\cite{TelescopeArray:2012qqu}.

Figure~\ref{fig:spectrum_features} shows the energy spectrum along with a fit using thrice-broken power laws. The black data points with error bars \textcolor{black}{show} the measured flux, while the red line indicates the best-fit model. The fit includes three break energies, $E_{\rm ankle}$, $E_{\rm shoulder}$, and $E_{\rm cutoff}$, as well as four spectral indices: $p_1$, $p_2$, $p_3$, and $p_4$. The ankle is located at $E_{\rm ankle} = 10^{18.70 \pm 0.01}$~eV, where the spectral index changes from $p_1 = -3.28 \pm 0.02$ to $p_2 = -2.62 \pm 0.03$. The softening feature, the shoulder, appears at $E_{\rm shoulder} = 10^{19.15 \pm 0.08}$~eV, with a subsequent index of $p_3 = -2.82 \pm 0.04$. The final break, the cutoff, occurs at $E_{\rm cutoff} = 10^{19.83 \pm 0.03}$~eV, where the spectrum steepens to $p_4 = -4.67 \pm 0.4$.

To evaluate the statistical significance of the cutoff at $10^{19.83}$~eV, we compare the expected and observed event counts above this energy. In the absence of a cutoff, 173.7 events are expected, while only 97 are observed, corresponding to a chance probability of $1.6 \times 10^{-10}$, or a $6.3\sigma$ significance.

A similar analysis for the shoulder at $10^{19.15}$~eV shows that 2156.4 events would be expected between $10^{19.15}$~eV and $10^{19.83}$~eV without the feature, whereas 1921 events are observed. This yields a chance probability of $1.3 \times 10^{-7}$, equivalent to a $5.2\sigma$ significance. This softening is consistent with the "instep" feature reported by the Auger at $10^{19.11 \pm 0.03}$~eV~\cite{PierreAuger:2020qqz}. After applying a +9\% energy rescaling~\cite{TelescopeArray:2021zox, SpectrumWG:2023wti} to the Auger energy scale, the spectral feature aligns with the TA shoulder at approximately $10^{19.15}$~eV, which demonstrates good agreement between observations in the northern and southern skies.

\vspace{-3mm}
\section{Declination Dependence in the Cosmic Ray Energy Spectrum}
\vspace{-3mm}

\begin{figure}[t]
    \centering
    \includegraphics[width=0.65\linewidth]{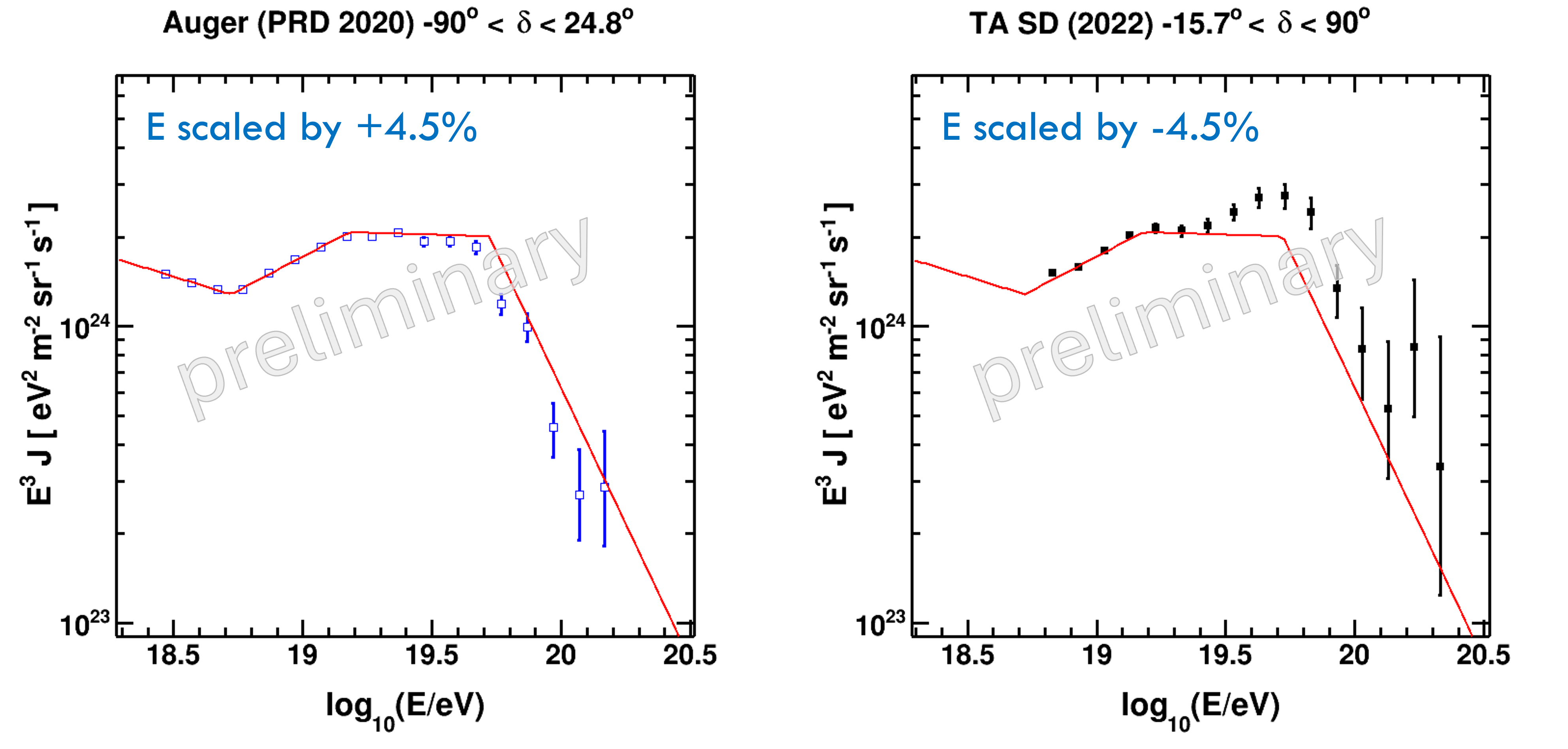}\\
    \includegraphics[width=0.65\linewidth]{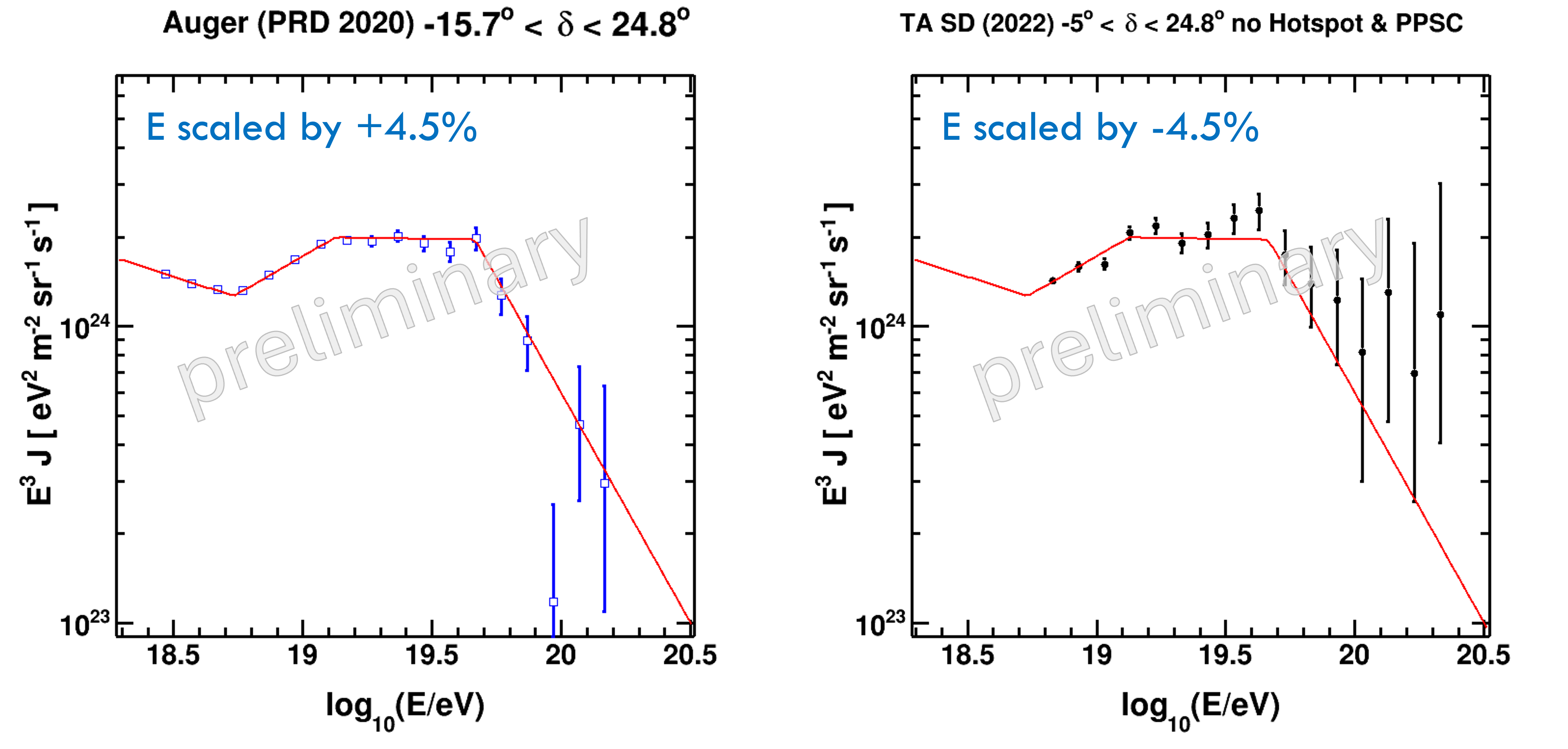}
    \caption{\textbf{Energy spectra measured by Auger and TA across their full apertures (top) and in the common sky with cuts applied to TA data (bottom).} The blue open squares represent Auger data with a $+4.5\%$ energy shift, and the black solid squares represent TA data with a $-4.5\%$ shift. The red lines in the top panels show the same broken power-law model obtained from a simultaneous fit to both spectra, yielding a significance of $8\sigma$. The bottom-left and bottom-right panels show Auger and TA data, respectively, in the common sky region with cuts applied to TA. The red lines in the bottom panels represent the result of the simultaneous fit, which yields a significance of $1.8\sigma$.}
    \label{fig:spec_auger_ta}
    \vspace{-3mm}
\end{figure}


In this section, we report on the observation of declination dependence in the cosmic ray energy spectrum. In 2017, a declination dependence in the energy spectrum measured by TA was first identified that shows a difference between lower and higher declination bands divided at $\delta = 24.8^\circ$~\cite{Ivanov:2017rwl}. We have recently revisited this phenomenon using an updated methodology~\cite{TelescopeArray:2024tbi}, examining the UHECR energy spectrum across both the northern and southern skies.

To evaluate the consistency between the TA and Auger energy spectra, we perform a simultaneous fit under the null hypothesis that both spectra originate from the same parent distribution. The simultaneous fit uses a thrice-broken power law model and is carried out using the binned log-likelihood method described in Eq. 40.16 of the Particle Data Group~\cite{ParticleDataGroup:2024cfk}.

The TA event selection criteria follow those utilized in the anisotropy studies and by the TA–Auger Joint Spectrum Working Group study for common declination analyses, described in~\cite{TelescopeArray:2025ani}.

The primary difference from the selection criteria described in Section~\ref{sec:spec_features} is the extended zenith angle cut to $55^\circ$, which increases sky coverage down to $\delta = -15.7^\circ$. To maintain reliable energy and angular resolution, a minimum energy threshold of $10^{18.8}$~eV is applied. These criteria are also used in anisotropy studies. This analysis includes 14 years of TA SD data with $E \geq 10^{18.8}$~eV and Auger data with $E \geq 10^{18.4}$~eV~\cite{PierreAuger:2020qqz, SpectrumWG:2023wti}.

The top panels of Figure~\ref{fig:spec_auger_ta} present the simultaneous fit results for both experiments across their full apertures. TA views the declination region $-15.7^\circ < \delta < +90^\circ$, while Auger observes $-90^\circ < \delta < +24.8^\circ$. Auger data (left panel) are shown as blue open squares after a $+4.5\%$ energy rescaling, while TA data (right panel) are shown as black solid squares after a $-4.5\%$ rescaling. These adjustments follow the Working Group’s findings, which demonstrated agreement in the ankle region~\cite{TelescopeArray:2021zox}. The red lines represent the same broken power law fit, yielding a total log-likelihood of 130.33 with 26 degrees of freedom. It corresponds to a Poisson probability of $7.5 \times 10^{-16}$. This result indicates an $8\sigma$ difference in the UHECR spectra between the northern and southern skies.

To validate this \textcolor{black}{methodology}, we examine the common sky region, $-15.7^\circ < \delta < +24.8^\circ$, \textcolor{black}{which is} observed by both experiments. \textcolor{black}{We introduce selection cuts to isolate potential sources of the apparent discrepancy for a more direct comparison within this common sky region. Specifically, we exclude} the rapidly declining TA exposure at its southernmost edge (below $\delta = -5^\circ$) \textcolor{black}{as well as} the excesses of events associated with medium-scale anisotropies such as the Hotspot and the Perseus-Pisces Supercluster (PPSC) excess~\cite{TelescopeArray:2014tsd, TelescopeArray:2021dfb, TelescopeArray:2025ykm, TelescopeArray:2025ani}, which extend into the common sky and lie near the northernmost edge of Auger’s exposure~\cite{TelescopeArray:2024tbi}.

The bottom panels of Figure~\ref{fig:spec_auger_ta} show the fit results in the common sky after applying these cuts to the TA data. The total log-likelihood is 40.12 with 26 degrees of freedom, corresponding to a Poisson probability of $3.8 \times 10^{-2}$ and a significance of $1.8\sigma$. This indicates no statistically significant difference between the two spectra in the common sky. This agreement reinforces the reliability of the simultaneous fit results obtained from the full-aperture spectra of both experiments, which indicate differences between the northern and southern skies in the UHECR energy spectrum.

\textcolor{black}{As a final step, we evaluate the significance while taking systematic uncertainty into account. Specifically, we rescale the energy by the $\pm 1\sigma$ uncertainty in the energy ratio between SD and FD energies, as shown by the gray band in the middle panel of Figure 1. Using this rescaled energy, we perform the simultaneous fit again. When the energy is increased, the result of the simultaneous fit is 10.0$\sigma$; when the energy is decreased, the result is 5.6$\sigma$. Taking the energy scale uncertainty into account, we evaluate the significance of the declination dependence to be $8.0^{+2.0}_{-2.4}\sigma$.}

\vspace{-3mm}
\section{Discussion}
\vspace{-3mm}

\begin{figure}[t]
    \centering
    \includegraphics[width=0.85\linewidth]{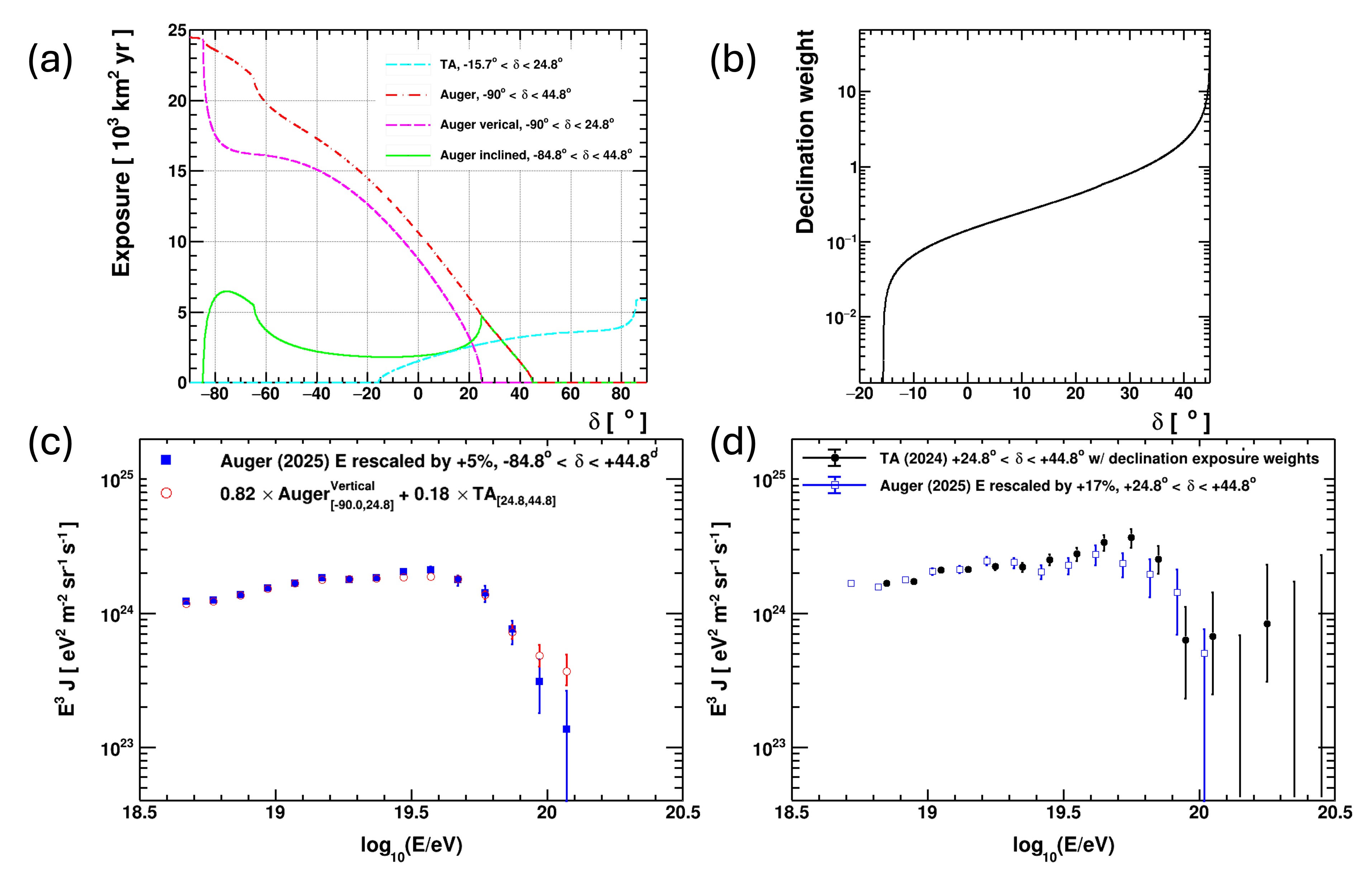}
    \caption{\textbf{Comparison of Auger and TA exposures and spectra} (a) Exposure distributions for the Auger vertical, Auger inclined, and TA spectra, highlighting the complementary coverage of the declination band. (b) Exposure ratio of TA to Auger as a function of declination, illustrating the mismatch in coverage. (c) Comparison of the Auger inclined spectrum and a weighted mix of the Auger vertical spectrum (up to $24.8^\circ$) and the TA spectrum ($24.8^\circ < \delta < 44.8^\circ$), showing agreement. (d) Comparison of the Auger inclined spectrum and the TA spectrum in the same band weighted by the exposure ratio, showing consistency between the two experiments in $24.8^\circ < \delta < 44.8^\circ$.
    }
    \label{fig:exposure_weight}
    \vspace{-6mm}
\end{figure}

A question arose at the Cosmic Ray Indirect spectrum session about the relevance of the coverage of the Auger inclined spectrum in the declination band of $24.8^\circ<\delta<44.8^\circ$. Figure~\ref{fig:exposure_weight}(a) shows the exposures of the Auger vertical spectrum, Auger inclined spectrum, and the TA spectrum. The inclined spectrum's exposure is concentrated in the lower part of the $\delta>24.8^\circ$ band, and the TA exposure is concentrated in the higher part of the band: a classic mismatch. Figure~\ref{fig:exposure_weight}(b) shows the ratio of the TA/Auger exposures as a function of declination. 

In order to understand the level of agreement between the two experiments, in Figure~\ref{fig:exposure_weight}(c), we plot the inclined spectrum (shown in~\cite{PierreAuger:2025hnw}) with a +5\% energy rescaling, plus a mixture of the vertical spectrum up to its limit of $24.8^\circ$ and TA spectrum from $24.8^\circ<\delta<44.8^\circ$, where they are weighted according to the fraction of the inclined spectrum's exposure each covers (82\% and 18\%, respectively). There is no disagreement. In Figure~\ref{fig:exposure_weight}(d), we plot the inclined spectrum from $24.8^\circ<\delta<44.8^\circ$ (strictly speaking, this is the extrapolation of the Auger combined spectrum), applying a +17\% energy rescaling. We also show the TA spectrum in the same declination band weighted by the ratio of Auger inclined/TA exposures. Again, there is no disagreement. \textcolor{black}{The shape of the Auger exposure greatly reduces its sensitivity to the spectral difference seen in the TA data.}


Weighting the TA events by the ratio of Auger/TA exposures as a function of declination also allows us to estimate how significant the TA Hotspot would be seen by the Auger. While the Hotspot is observed with a local Li-Ma significance of approximately $5\sigma$ by TA, it would appear with about $2\sigma$ local significance to Auger, i.e., statistically insignificant~\cite{TelescopeArray:2025ani}.

\vspace{-3mm}
\section{Summary}
\vspace{-3mm}

Over 16 years, the Telescope Array has operated with exceptional stability, enabling the collection of high-quality data. This consistency has allowed for rigorous validation of our Monte Carlo simulation framework through direct comparison with observations, ensuring reliable simulations for UHECR studies, including event reconstruction and detector exposure.

The robustness of the TA SD energy reconstruction is confirmed by three independent methods: (1) consistency between thrown and reconstructed MC energies, (2) comparisons between SD and FD energies, and (3) agreement between CIC-derived and standard TA energies. These results affirm the linearity and reliability of the TA SD energy scale. 

We report key spectral features in the 16-year TA SD dataset: the ankle at $10^{18.70 \pm 0.01}$~eV, the shoulder at $10^{19.15 \pm 0.08}$~eV, and the high-energy cutoff at $10^{19.83 \pm 0.03}$~eV, with statistical significances of $5.2\sigma$ and $6.3\sigma$ for the shoulder and cutoff, respectively. These features offer critical insights into UHECR origin and propagation.

A significant spectral difference between the northern and southern hemispheres is observed, with an $8.0\sigma$ discrepancy from a joint TA–Auger fit. However, in the common sky region, this difference is reduced to $1.8\sigma$ after applying cuts, indicating no statistically significant difference in the common sky. Taking the energy scale uncertainty into account, we evaluate the $8.0\sigma$ discrepancy and obtain $8^{+2.0}_{-2.4}\sigma$. These findings suggest that the observed spectral difference originates from a genuine variation in UHECR flux between the northern and southern skies.

\newpage
\section*{Full Authors List: Telescope Array Collaboration}

\makeatletter
\newcommand{\ssymbol}[1]{^{\@fnsymbol{#1}}}
\makeatother
\par\noindent
R.U.~Abbasi$^{1}$,
T.~Abu-Zayyad$^{1,2}$,
M.~Allen$^{2}$,
J.W.~Belz$^{2}$,
D.R.~Bergman$^{2}$,
F.~Bradfield$^{3}$,
I.~Buckland$^{2}$,
W.~Campbell$^{2}$,
B.G.~Cheon$^{4}$,
K.~Endo$^{3}$
% ...
\bigskip
\par\noindent
{\footnotesize\it
$^{1}$ Department of Physics, Loyola University-Chicago, Chicago, Illinois 60660, USA \\
$^{2}$ High Energy Astrophysics Institute and Department of Physics and Astronomy, University of Utah, Salt Lake City, Utah 84112-0830, USA \\
$^{3}$ Graduate School of Science, Osaka Metropolitan University, Sugimoto, Sumiyoshi, Osaka 558-8585, Japan \\
$^{4}$ Department of Physics and The Research Institute of Natural Science, Hanyang University, Seongdong-gu, Seoul 426-791, Korea \\
% ...

\let\thefootnote\relax\footnote{$\ssymbol{1}$ Presently at: Purple Mountain Observatory, Nanjing 210023, China}
\let\thefootnote\relax\footnote{$\ssymbol{2}$ Presently at: Physics Department, Brookhaven National Laboratory, Upton, NY 11973, USA}
\let\thefootnote\relax\footnote{$\ssymbol{3}$ Presently at: Korea Institute of Geoscience and Mineral Resources, Daejeon, 34132, Korea}
\let\thefootnote\relax\footnote{$\ssymbol{4}$ Presently at: NASA Marshall Space Flight Center, Huntsville, Alabama 35812, USA}
\addtocounter{footnote}{-1}\let\thefootnote\svthefootnote
}
\par\noindent

\section*{Acknowledgements}

The Telescope Array experiment is supported by the Japan Society for
the Promotion of Science(JSPS) through
Grants-in-Aid
for Priority Area
431,
for Specially Promoted Research
JP21000002,
for Scientific  Research (S)
JP19104006,
for Specially Promoted Research
JP15H05693,
for Scientific  Research (S)
JP19H05607,
for Scientific  Research (S)
JP15H05741,
for Science Research (A)
JP18H03705,
for Young Scientists (A)
JPH26707011,
for Transformative Research Areas (A)
JP25H01294,
for International Collaborative Research
24KK0064,
and for Fostering Joint International Research (B)
JP19KK0074,
by the joint research program of the Institute for Cosmic Ray Research (ICRR), The University of Tokyo;
by the Pioneering Program of RIKEN for the Evolution of Matter in the Universe (r-EMU);
by the U.S. National Science Foundation awards
PHY-1806797, PHY-2012934, PHY-2112904, PHY-2209583, PHY-2209584, and PHY-2310163, as well as AGS-1613260, AGS-1844306, and AGS-2112709;
by the National Research Foundation of Korea
(2017K1A4A3015188, 2020R1A2C1008230, and RS-2025-00556637) ;
by the Ministry of Science and Higher Education of the Russian Federation under the contract 075-15-2024-541, IISN project No. 4.4501.18, by the Belgian Science Policy under IUAP VII/37 (ULB), by National Science Centre in Poland grant 2020/37/B/ST9/01821, by the European Union and Czech Ministry of Education, Youth and Sports through the FORTE project No. CZ.02.01.01/00/22\_008/0004632, and by the Simons Foundation (MP-SCMPS-00001470, NG). This work was partially supported by the grants of the joint research program of the Institute for Space-Earth Environmental Research, Nagoya University and Inter-University Research Program of the Institute for Cosmic Ray Research of University of Tokyo. The foundations of Dr. Ezekiel R. and Edna Wattis Dumke, Willard L. Eccles, and George S. and Dolores Dor\'e Eccles all helped with generous donations. The State of Utah supported the project through its Economic Development Board, and the University of Utah through the Office of the Vice President for Research. The experimental site became available through the cooperation of the Utah School and Institutional Trust Lands Administration (SITLA), U.S. Bureau of Land Management (BLM), and the U.S. Air Force. We appreciate the assistance of the State of Utah and Fillmore offices of the BLM in crafting the Plan of Development for the site.  We thank Patrick A.~Shea who assisted the collaboration with much valuable advice and provided support for the collaboration’s efforts. The people and the officials of Millard County, Utah have been a source of steadfast and warm support for our work which we greatly appreciate. We are indebted to the Millard County Road Department for their efforts to maintain and clear the roads which get us to our sites. We gratefully acknowledge the contribution from the technical staffs of our home institutions. An allocation of computing resources from the Center for High Performance Computing at the University of Utah as well as the Academia Sinica Grid Computing Center (ASGC) is gratefully acknowledged.

\end{document}